\documentclass[manuscript,screen,nonacm]{acmart}

\renewcommand\footnotetextcopyrightpermission[1]{}
\setcopyright{none}

\usepackage{algorithmic}
\usepackage{multirow}
\usepackage{array}
\graphicspath{{figures/}}
\begin{document}

\title{Cross-Layer Analysis of Thermal Tuning Stalls in Wafer-Scale Optical Interconnects for LLM MoE Training}

\author{Seongwon Yoon}
\affiliation{%
  \institution{Georgia Institute of Technology}
  \city{Atlanta}
  \state{GA}
  \country{USA}}
\email{gabrielyoon@gatech.edu}

\author{Pin-Jun Chen}
\affiliation{%
  \institution{Georgia Institute of Technology}
  \city{Atlanta}
  \state{GA}
  \country{USA}}
\email{pinjun.chen@gatech.edu}

\author{Shimeng Yu}
\affiliation{%
  \institution{Georgia Institute of Technology}
  \city{Atlanta}
  \state{GA}
  \country{USA}}
\email{shimeng.yu@ece.gatech.edu}

\begin{abstract}
Mixture-of-experts (MoE) training depends heavily on all-to-all communication, which wafer-scale optical interconnects with dense wavelength-division multiplexing can serve. Their microring resonators rely on thermo-optic tuning, while MoE compute bursts swing the photonic-layer temperature by about 10~K within tens of milliseconds. This paper quantifies the communication stall by coupling packet-level network simulation, transient thermal simulation of a 3D-stacked GPU and photonic die, and a ring detuning criterion, and by feeding the stall back into the network timeline until both agree. Including this feedback, a tracking loop at the measured 5~nm/s lengthens the training iteration of Mixtral 8x7B and LLaMA-MoE 6.7B by factors of about 1.16 and 1.25 at full model depth. Measured H100 die temperatures match the modeled swing of 300~ms bursts within 15\%. A loop slewing eight times faster, a heater driven at each kernel launch, or a dummy load near 75\% of peak power removes the stall. An athermalized lithium-niobate ring with a non-volatile ferroelectric setpoint removes it with no holding power or fast loop, which makes it suitable for wafer-scale optical interconnects.
\end{abstract}

\ccsdesc[500]{Hardware~Emerging optical and photonic technologies}
\ccsdesc[500]{Hardware~Photonic and optical interconnect}
\ccsdesc[300]{Hardware~Temperature simulation and estimation}
\ccsdesc[300]{Computer systems organization~Interconnection architectures}
\ccsdesc[100]{Computing methodologies~Distributed computing methodologies}

\keywords{mixture-of-experts, optical interconnects, wafer-scale systems, microring resonators, thermo-optic tuning, cross-layer co-simulation}

\maketitle

\section{Introduction}

The rapid growth of large-scale artificial intelligence (AI) models has substantially increased the communication demands placed on modern computing systems. Large language models (LLMs) are trained on distributed accelerator platforms whose performance is increasingly limited by data movement as well as computation. The trend is amplified in mixture-of-experts (MoE) models, which add expert parallelism to data, tensor, and pipeline parallelism and therefore generate intensive all-to-all traffic, making the interconnect fabric a first-order design constraint.

Unlike dense transformers, MoE models route each token to a few selected experts, so their communication is bursty and non-uniform. Expert communication occupies a large fraction of the iteration time, and skewed expert selection concentrates traffic and computation on a subset of accelerators. MoE performance is therefore sensitive to link bandwidth, latency, and any overhead that delays a communication round.

The pressure is greatest in the scale-up domain, where accelerators within a node or package must exchange data at high throughput and low latency. Electrical interconnects face growing constraints in bandwidth density, energy efficiency, and reach as systems grow. Optical interconnects have emerged as an alternative, as studies of wafer-scale silicon photonics \cite{lightmatter} and optical interposer bridges for AI accelerators \cite{celestialai} show.

A key advantage of optical interconnects is dense wavelength-division multiplexing (DWDM), which carries many wavelength channels on one waveguide \cite{dwdm} and scales bandwidth without a proportional increase in the number of physical links. Packaging platforms such as TSMC's COUPE place photonic interfaces directly next to processors \cite{tsmc_coupe}, and system-on-wafer integration places up to 16 full-reticle compute dies and their HBM on one reconstructed wafer \cite{tsmc_sow}.

Bandwidth alone does not settle the question. DWDM links select, modulate, and route wavelengths with microring resonators (MRRs), which must be held on resonance, typically by a heater under closed-loop control \cite{mrr}. MRRs are sensitive to temperature \cite{imec_Thermal}, and in a wafer-scale system the photonic layer sits directly beneath GPUs whose power swings by hundreds of watts between compute bursts and idle intervals. Whenever the ring drifts faster than its tuning loop can follow, the link stops carrying data until the loop catches up, and a fabric with ample nominal bandwidth can still deliver its data late.

How much this matters has not been quantified for real training workloads. Prior studies characterize thermal transients at the device level or use analytical workload models, and none couples these transients to the timing of the communication rounds, which determines whether a drift actually delays communication. We therefore close the loop between the layers. The per-device timelines of a packet-level network simulation drive a transient thermal model of the stack, the resonance drift sets a per-round communication stall, and the stalls are fed back into the network simulation until the timeline and the stall agree. We cross-check the thermal model against measurements on a real GPU, and we translate the result into what a tuning loop, a predictive heater drive, a dummy load, or a drift-free ring would each have to deliver.

The main contributions of this paper are as follows.
\begin{itemize}
\item \textbf{A closed-loop cross-layer framework, cross-checked on hardware.} A packet-level network simulation, a transient thermal model grounded in a published characterization of 2.5D and 3D co-packaged optics \cite{imec_TCPMT}, and an optical-domain stall criterion are chained so that every delay is derived rather than assumed, and the chain is solved self-consistently. Die-temperature telemetry of an H100 GPU under synthetic and real MoE workloads matches the modeled swing of 300~ms bursts within 15\%, confirms the 700~W power level of the model, and exposes a second, package-level drift timescale.
\item \textbf{The tuning stall is a first-order, thermally coupled effect.} On a wafer fabric whose stall-free iteration time is within 3\% and 16\% of that of an ideal fat-tree for Mixtral 8$\times$7B and LLaMA-MoE 6.7B, a tracking loop at the measured 5~nm/s lengthens the training iteration at full model depth by factors of about 1.16 and 1.25 once the stall is fed back. In the simulated configuration the factors are 1.32 and 1.55 with the feedback and 1.58 and 2.82 in an open-loop estimate. Capturing it requires the thermal feedback, because the stall reshapes the duty cycle of each GPU and thereby its own thermal cause. Each collective waits for the participating device whose rounds fall in that device's cooling transients.
\item \textbf{Quantified requirements for avoiding the stall.} A loop slewing at 40~nm/s or reacting within 0.5~ms, a heater pre-driven within about 1~ms of each GPU kernel launch, or a dummy load above 75\% of peak power each removes the stall, and added thermal mass does not. The fabric itself needs a co-packaged-optics (CPO) port on every reticle rather than a gateway shared between wafers.
\item \textbf{An athermal ring with a non-volatile ferroelectric setpoint as the least costly mitigation.} Compared with the other options, it is the only one that removes the stall with no holding power, no fast loop, no signal from the GPU, and no dependence on the workload. We quantify how close to the operating temperature its athermal point must lie and show why a ferroelectric actuator cannot instead track the drift.
\end{itemize}

\section{Background}
\label{sec:background}

\subsection{MoE Training and All-to-All Communication}

An MoE block replaces the feed-forward network (FFN) of a transformer with a set of parallel experts. After attention, a gate computes routing scores and selects a few experts for each token, the selected experts process the token, and their outputs are combined into the block output. Only the selected experts run, so the model capacity grows while the computation per token stays close to that of a much smaller dense model.

\begin{figure}[t]
\centering
\includegraphics[width=\linewidth]{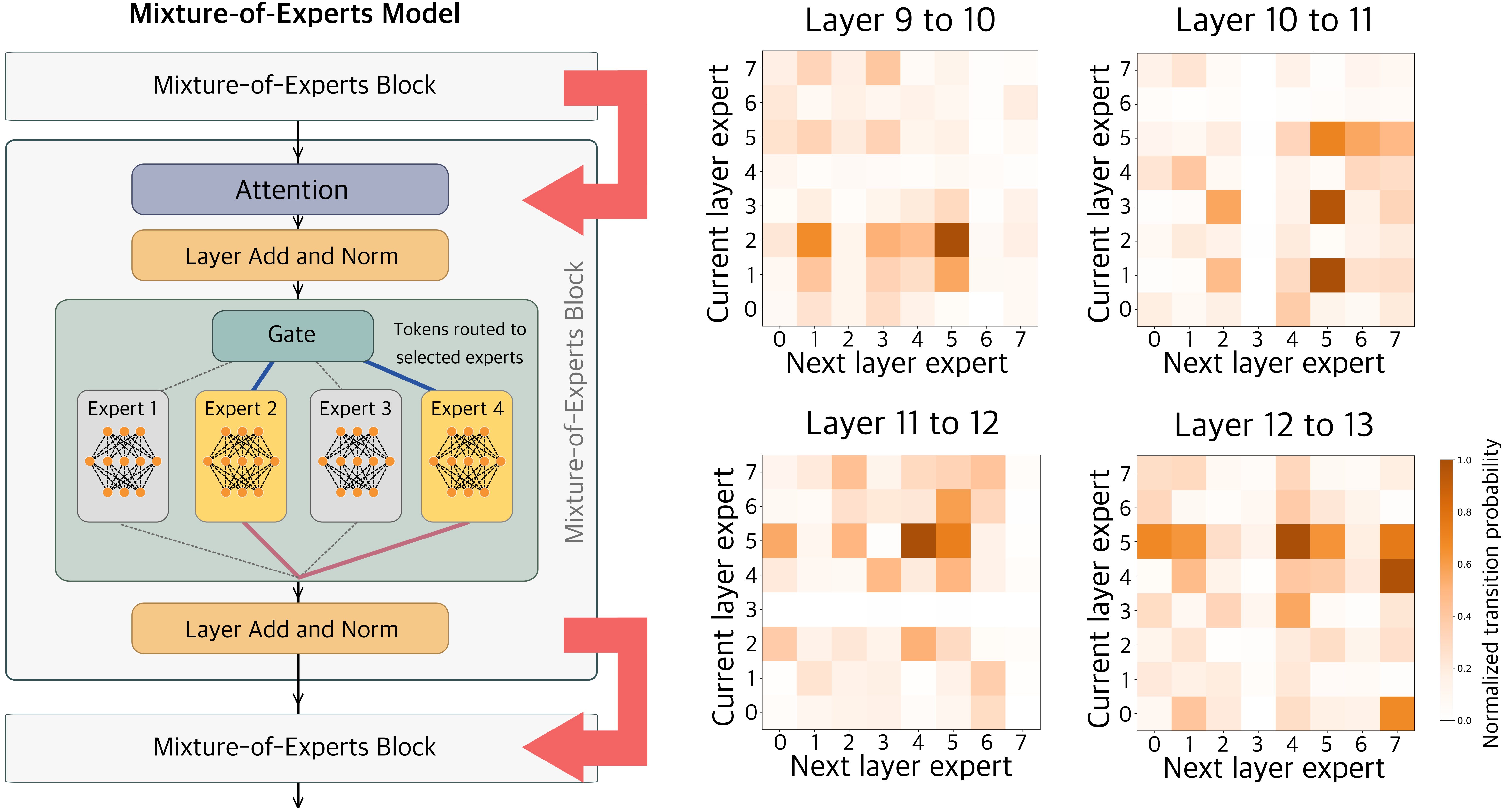}
\caption{Overview of an MoE block and expert-selection behavior in Mixtral 8$\times$7B. The left diagram illustrates the computation flow within an MoE layer, including gating, expert routing, and aggregation. The right heatmaps show normalized transition probabilities from the expert chosen at one layer to the expert chosen at the next, for layers 9 to 13. A few transitions dominate, indicating non-uniform and persistent routing.}
\label{fig:expert_transition}
\end{figure}

Expert selection is not uniform in practice \cite{chronos}. Fig.~\ref{fig:expert_transition} shows layer-to-layer expert transition probabilities for Mixtral 8$\times$7B. A few transitions dominate, so some experts, and the devices that host them, receive a disproportionate share of the tokens.

MoE training combines data parallelism (DP) \cite{dp}, tensor parallelism (TP) \cite{tp}, and pipeline parallelism (PP) \cite{pp} with expert parallelism (EP) \cite{gshard}, which places different experts on different accelerators. Each MoE block then needs two all-to-all phases. Token dispatch sends token representations to the devices that host the selected experts, and token combine returns the expert outputs. The resulting traffic varies over time, is non-uniform across experts, and concentrates on a subset of GPU pairs, so MoE training is sensitive to interconnect bandwidth, latency, and any per-round overhead.

In our measurement of a Mixtral-style MoE model trained on eight H100 GPUs (Section~\ref{sec:thermal_grounding}), the dispatch and combine phases occupy 44\% of the iteration on average and up to 57\%. In the simulated wafer-scale configurations of this paper, each device spends about half of the iteration waiting on all-to-all rounds or on the pipeline rather than computing, and this alternation drives its temperature swing.

\subsection{Optical Scale-Up Interconnects and Microring Resonators}

This traffic stresses the scale-up domain, where accelerators within a node or package exchange data at high throughput and low latency. Electrical I/O is limited by the beachfront, the usable perimeter, of a chip \cite{celestialai}, whereas compute capability scales with its area. Optical interconnects relax this limit. In co-packaged optics (CPO), the optics are brought into the electronic package of the processor to shorten the electrical paths \cite{corning}. In optical interposers, photonic components sit beneath the compute and memory chiplets and remove the beachfront limit by placing photonic elements at the point of compute \cite{celestialai}, so bandwidth scales with chip area rather than chip perimeter. Dense wavelength-division multiplexing (DWDM) further multiplies the bandwidth of each waveguide. A five-channel microring modulator array has carried 1~Tb/s on one bus waveguide, and a single comb laser supplies more than 50 usable wavelength lines \cite{dwdm}.

Mach-Zehnder interferometers (MZIs), microring resonators (MRRs), and electro-absorption modulators (EAMs) offer different trade-offs in footprint, thermal robustness, power, and optical bandwidth \cite{witzens, celestialAItrx}. MZIs are broadband and thermally robust but long \cite{witzens}, EAMs are compact and broadband \cite{celestialAItrx}, and MRRs are compact and wavelength selective, so each ring acts on one channel of a DWDM comb. This selectivity makes MRRs the natural device for DWDM links with 32 wavelengths per waveguide, and it also makes them sensitive to temperature. The resonance of a silicon MRR shifts strongly with temperature \cite{padmaraju_thermal}, so the thermal fluctuations caused by computation in tightly integrated accelerator systems shift the resonances and require continuous retuning. DWDM links may therefore be limited not only by their nominal bandwidth but also by how quickly their rings can be kept aligned.

\section{Wafer-Scale Photonic System Architecture}

\begin{figure}[t]
\centering
\includegraphics[width=0.75\linewidth]{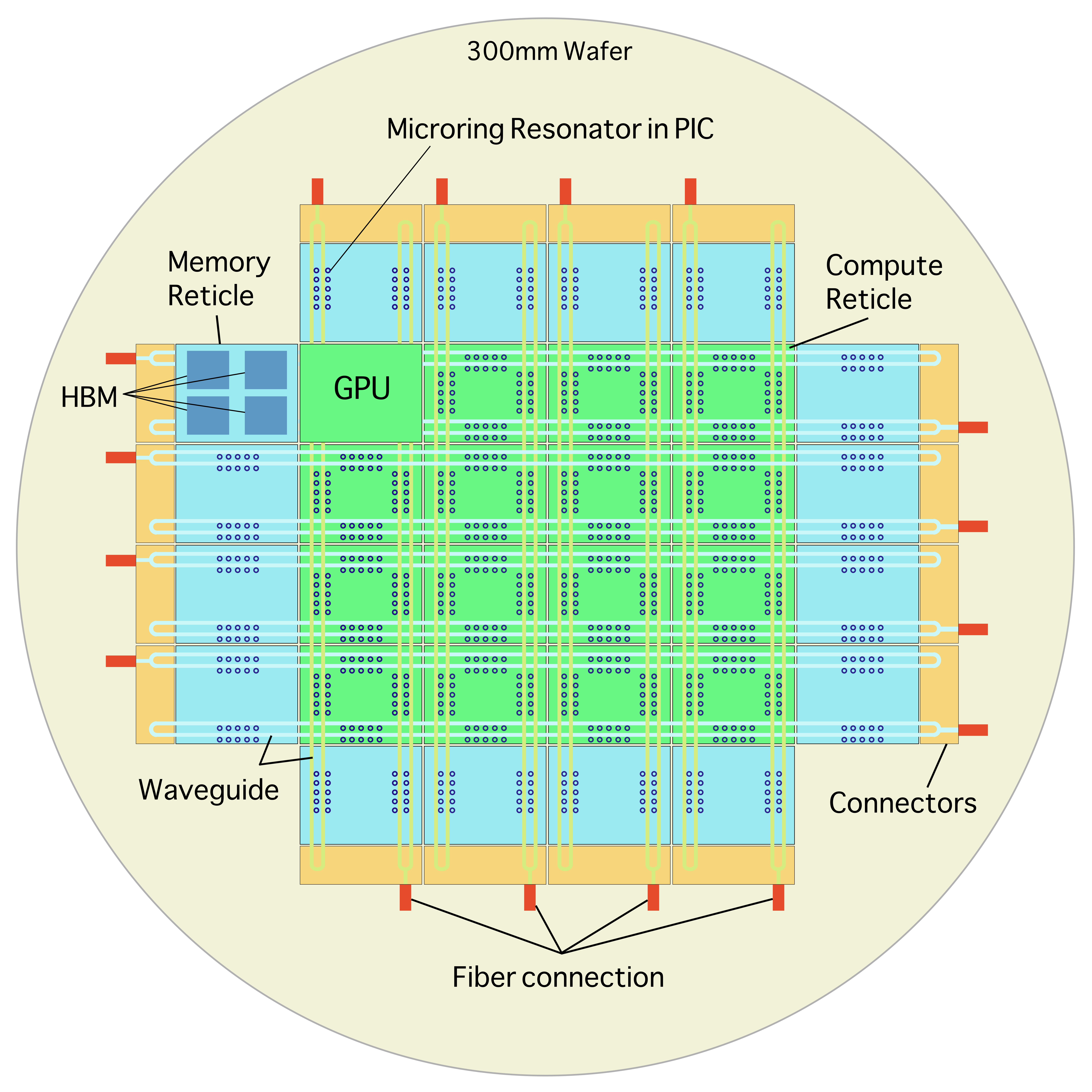}
\caption{Overview of the wafer-scale photonic system architecture. A 300~mm wafer integrates compute reticles (GPU chiplets) near the center and memory reticles (HBM) around the periphery. The photonic interconnect layer uses a grid of SiN waveguides for global communication, while microring resonators (MRRs) in the PIC layer provide wavelength-selective coupling and modulation. Off-wafer fibers are shown schematically; at the design point each reticle carries its own CPO port (Section~\ref{sec:wsc_sin}).}
\label{fig:wsc_arch}
\end{figure}

As shown in Fig.~\ref{fig:wsc_arch}, the architecture distributes compute and memory reticles across a wafer-scale photonic fabric that provides high-bandwidth optical communication throughout the system.

\subsection{3D Integration of Compute and Memory Reticles}

\begin{figure}[t]
\centering
\includegraphics[width=\linewidth]{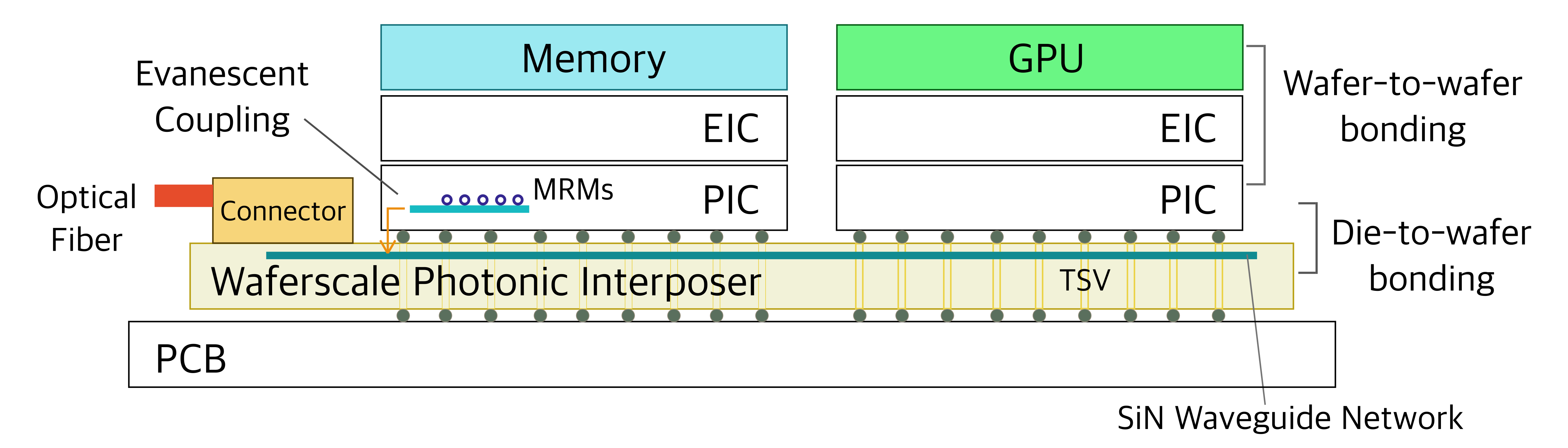}
\caption{Cross-sectional view of the 3D-integrated optical-interposer system. GPU and memory dies are vertically integrated with EIC and PIC layers through wafer-to-wafer bonding and then attached to a wafer-scale photonic interposer through die-to-wafer bonding. Optical signals couple into the SiN waveguide network through vertical evanescent couplers, while MRRs in the PIC provide wavelength-selective operation.}
\label{fig:wsc_3dstack}
\end{figure}

The system uses 3D-integrated compute and memory reticles (Fig.~\ref{fig:wsc_3dstack}) that combine electronic and photonic components. Each reticle is formed in two bonding steps. The compute or memory die and its electrical integrated circuit (EIC) are first integrated with a photonic integrated circuit (PIC) through wafer-to-wafer (W2W) hybrid bonding, and the resulting stacks are then attached to the wafer-scale optical interposer by collective die-to-wafer (D2W) bonding, in which the dies are placed on a carrier and transferred to the target wafer in one bonding step \cite{imec_D2W}. This two-step scheme gives dense local electrical integration while connecting each reticle to the global optical fabric.

The top layer contains the GPU or memory die. Beneath it, the EIC implements the electrical front end for optical communication, including serializer and deserializer (SerDes) blocks, modulator drivers, transimpedance amplifiers (TIAs), equalizers, and link management circuits. The PIC implements the optical functions of the reticle, including waveguides, modulators, photodetectors, and wavelength-selective microring resonators \cite{pic_3d}. The PIC couples to passive silicon nitride (SiN) waveguides in the interposer through vertical evanescent couplers based on inverse tapers \cite{coupler, imec_D2W}.

\subsection{Wafer-Scale SiN Photonic Interconnect}
\label{sec:wsc_sin}

The reticles communicate through a wafer-scale SiN photonic interposer spanning a 300~mm wafer. The waveguide platform is based on reticle-stitched SiN routing \cite{imec_WG}, in which long waveguide sections are patterned over multiple reticles and stitched with high alignment precision, giving continuous optical paths of several tens of centimeters with low propagation loss and negligible stitching penalty.

To support many waveguide crossings, the architecture adopts a multilayer SiN network \cite{wg_network, panelscale} in which orthogonal routing directions are assigned to different SiN layers, with high-precision SiO$_2$ planarization enabling vertical transitions and crossings. The interconnect is thus organized as an orthogonal grid of row and column waveguides. Reticle-to-reticle communication uses optical-electrical-optical (OEO) relay rather than on-wafer optical switching. At an intermediate reticle, the local EIC converts the incoming optical signal to the electrical domain, and the local PIC retransmits it into the next waveguide segment. This avoids on-wafer optical switches and regenerates the signal.

Following prior photonic-interposer assumptions, we use 32~Gb/s per wavelength and 32 wavelengths per waveguide, i.e., 1.024~Tb/s per waveguide \cite{panelscale}. To reach 1.5~TB/s (12~Tb/s) of die-to-die bandwidth per direction, the required number of wavelength channels is
\[
N_{\lambda}=\left\lceil \frac{12\ \mathrm{Tb/s}}{32\ \mathrm{Gb/s}} \right\rceil = 375,
\]
and with 32 wavelengths per waveguide the required number of waveguides is
\[
N_{\mathrm{WG}}=\left\lceil \frac{375}{32} \right\rceil = 12,
\]
which provides 12.288~Tb/s. Each wavelength needs one transmitter-side and one receiver-side resonator, so the endpoint interface uses 375 of each, or 750 in total (768 when all $12\times32$ wavelength slots are provisioned).

The architecture also provides optical interfaces for off-wafer connectivity. Each reticle stack carries a co-packaged-optics (CPO) port whose fibers leave the wafer through integrated connectors, so that traffic between wafers enters and leaves the fabric at the reticle that sources or sinks it rather than through a shared gateway. We model the off-wafer network that joins these ports as non-blocking, and Section~\ref{sec:baseline_topo} shows that per-reticle ports, rather than a shared gateway, let the fabric keep pace with a non-blocking electrical network.

\subsection{Photonic Tuning Mechanisms}
\label{sec:background_ferro}

\begin{figure}[t]
\centering
\includegraphics[width=0.5\linewidth]{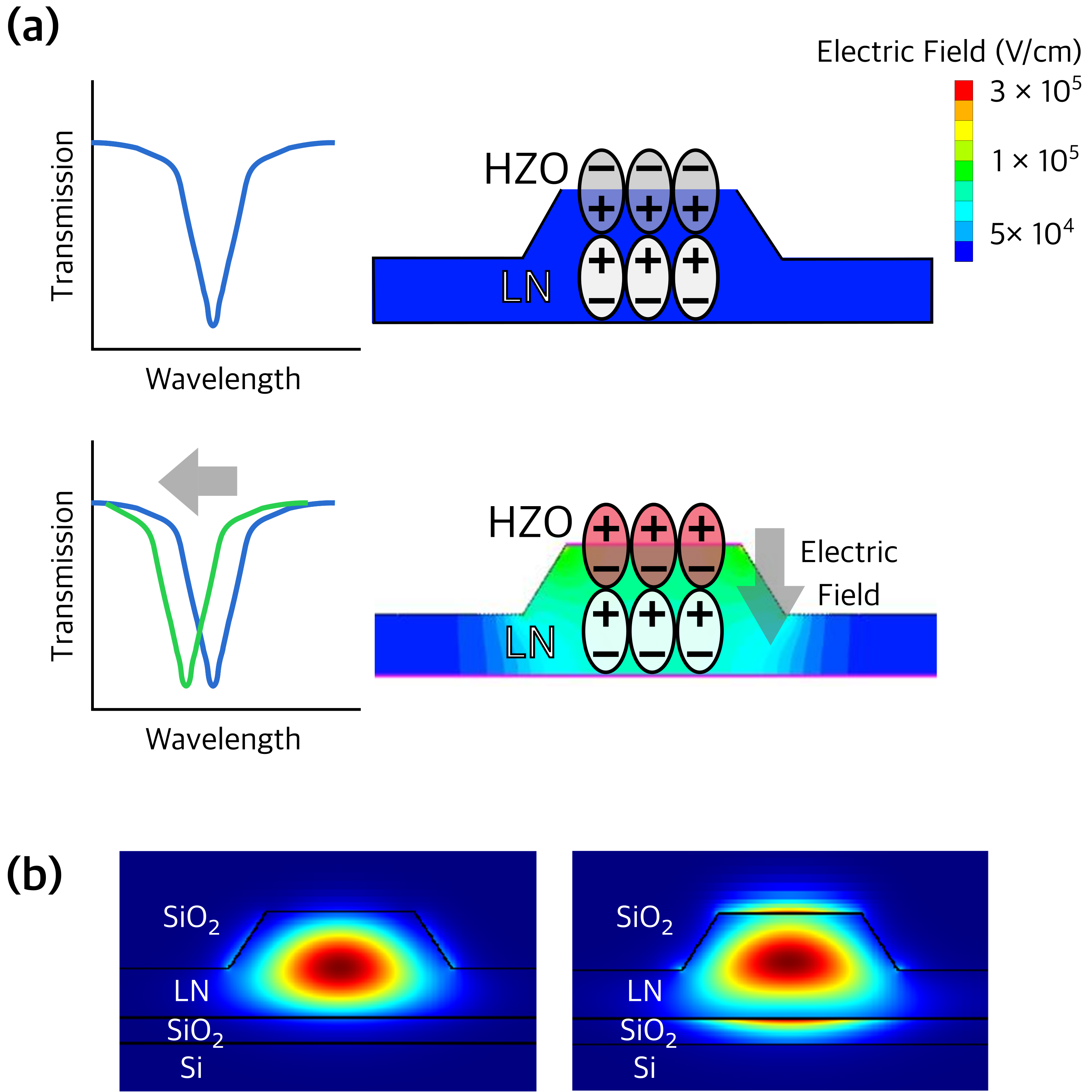}
\caption{Device-level modeling of a ferroelectric LiNbO$_3$ (LN) waveguide with an HZO gate stack. (a) Polarization switching under an applied electric field induces a non-volatile refractive-index change in the LN waveguide through the Pockels effect, shifting the transmission spectrum. (b) TE and TM mode profiles of the LN rib waveguide at the tuned wavelength, obtained in Ansys Lumerical using the TCAD-derived refractive-index distribution.}
\label{fig:fe_device}
\end{figure}

In conventional silicon photonics, MRRs are aligned by thermo-optic tuning, in which a heater changes the refractive index of the ring. A control loop in the EIC monitors the ring and adjusts the heater, and its speed is bounded by the signal-to-noise ratio and update rate of its monitor rather than by the heater itself, whose time constant is microseconds. A step of the compute-die (XPU) power from 0 to 100\% produces a thermal gradient of 1.78~K/ms, which for a 10~pm stabilization accuracy implies a controller reaction time of 84~$\mu$s \cite{imec_Thermal}, whereas an integrated drop-port locking loop reports a tracking speed of 5~nm/s (0.0625~K/ms) \cite{PID}. A heater can only add heat, so the ring must be biased above its coldest operating point by the full expected swing.

An alternative actuator is a lithium niobate (LiNbO$_3$, LN) ring with a hafnium zirconium oxide (HZO) ferroelectric gate stack \cite{ferroMRR}. An applied field changes the LN refractive index through the Pockels effect, and when the field exceeds the coercive field of HZO the polarization switches and remains after the field is removed, so the resonance offset persists without holding power \cite{ferroMRR}. With one ferroelectric gate, the reported device shifts its resonance by about 60~pm in six non-volatile states. With three gates placed around the ring, it shifts by about 120~pm in 16 states. In the single-gate device a state is written with a 10~$\mu$s pulse at 65~fJ and reset at 358~fJ, and the device has been cycled more than $10^7$ times \cite{ferroMRR}. The HZO film loses less than 15\% of its polarization below 90~$^\circ$C \cite{ferroMRR}, and the characteristic switching time of a 15~nm HZO film is 5.4~ns \cite{hzo_switching_speed}. We used technology computer-aided design (TCAD) to obtain the refractive-index distribution induced by the field in the HZO-gated LN waveguide and imported it into Ansys Lumerical to confirm that the TE and TM modes near 1550~nm remain guided under the applied field (Fig.~\ref{fig:fe_device}). A bare LN ring drifts by roughly 18 to 30~pm/K, from the 18~pm/K measured on a bare x-cut thin-film LN microring \cite{ln_athermal_broadband} to the value that the bulk extraordinary thermo-optic coefficient of LN near 60~$^\circ$C implies for a mode polarized along the extraordinary axis \cite{ln_toc_bulk}. Cladding it with TiO$_2$, whose thermo-optic coefficient is negative, cancels the first-order temperature dependence at a design temperature. The measured residual stays within $\pm$1~pm/K over 1480 to 1630~nm \cite{ln_athermal_broadband}, and the second-order coefficient is 0.37~pm/K$^2$ with the athermal point at room temperature \cite{ln_athermal}. Section~\ref{sec:endurance} examines what these two devices can and cannot do against the drift quantified below.

\section{Cross-Layer Thermal-to-Network Evaluation}
\label{sec:eval}

\subsection{Evaluation Setup}
\label{sec:experimental_setup}

\begin{table}[t]
\centering
\small
\caption{LLM MoE training configurations used in the network and thermal simulations.}
\label{tab:moe_config}
\renewcommand{\arraystretch}{1.15}
\begin{tabular}{|l|c|c|}
\hline
\textbf{Configuration} & \textbf{Mixtral 8$\times$7B} & \textbf{LLaMA-MoE 6.7B} \\
\hline
MoE blocks in the full model & 32 & 32 \\
\hline
MoE blocks per pipeline stage in the simulation & 1 & 1 \\
\hline
Number of experts & 8 & 16 \\
\hline
EP degree & 8 & 16 \\
\hline
TP degree & 4 & 1 \\
\hline
PP degree & 4 & 4 \\
\hline
DP degree & 2 & 2 \\
\hline
GPUs (wafers of 16) & 256 (16) & 128 (8) \\
\hline
GPUs per expert-parallel group (wafers spanned) & 32 (2) & 16 (1) \\
\hline
Sequence length & 4096 & 4096 \\
\hline
Micro-batches per iteration and data-parallel replica & 4 & 4 \\
\hline
\end{tabular}
\end{table}

\textbf{Workload and network model.} Mixtral 8$\times$7B \cite{mixtral8x7b}, with eight experts and top-2 routing, and LLaMA-MoE 6.7B are profiled on an NVIDIA H100 SXM 80~GB GPU. LLaMA-MoE 6.7B denotes the released LLaMA-MoE-3.0B (2/16) model, which has 16 experts with top-2 routing, 3.0B activated parameters, and 6.7B total parameters \cite{llama}. Profiling uses the FlexFlow \cite{flexflow} version in the MixNet artifact \cite{mixnet}, whose task-graph export records the forward and backward cost of every operator (attention, gating, expert feed-forward, all-to-all dispatch and combine, add-and-normalization) and exports the dataflow graph of one training iteration across all pipeline, expert-parallel, and data-parallel shards. The same graph drives the packet-level network simulation in the MixNet extension of htsim \cite{htsim, mixnet}, which routes tokens according to the measured expert-selection matrix. The per-device compute timelines produced by this simulation in turn drive the transient thermal analysis. Table~\ref{tab:moe_config} lists the parallelization settings. Each expert-parallel group spans two wafers for Mixtral and one for LLaMA-MoE, which matters because all-to-all traffic that leaves a wafer must cross the off-wafer ports. One correction is applied to the exported graphs. FlexFlow assigns zero bytes to the dispatch all-to-all (\texttt{Group\_by}) operators, so we size each dispatch round like the combine round of the same layer, micro-batch, and direction, since both move the same token payload; the H100 measurement of Section~\ref{sec:thermal_grounding} confirms equal byte counts in both directions.

\textbf{Wafer fabric parameters.} The simulated wafer implements Section~\ref{sec:wsc_sin} with a $4\times4$ array of GPU reticles per 300~mm wafer, an orthogonal row and column SiN waveguide grid, and OEO relay at row and column intersections. Each GPU owns 12 waveguides of 32 wavelengths at 32~Gb/s, i.e., 1.5~TB/s of egress. It has three row peers and three column peers, and each of these six links uses two waveguides, or 256~GB/s. Off-wafer connectivity uses one CPO port of 800~GB/s per reticle (Section~\ref{sec:baseline_topo}). A non-blocking fat-tree and a direct all-to-all (``flat'') fabric serve as baselines at the same 1.5~TB/s per-GPU injection bandwidth.

\textbf{GPU power model.} Each compute phase of the per-device timeline receives a power level set by its arithmetic intensity on the H100 SXM roofline \cite{nvidia_h100}. Attention and expert feed-forward general matrix multiplications (GEMMs), whose intensity is of the order of the sequence length, are compute-bound and receive the full 700~W XPU power. The gate GEMM, whose intensity is of the order of half the expert count, receives 40\%, elementwise add-and-normalization receives 20\%, and any interval without a kernel, spent waiting on an all-to-all round or on the pipeline, receives 12\%. The 700~W level matches the 692~W that the H100 draws under sustained GEMM load in our measurement and the 692~W at which the per-GPU power of H200 prefill instances saturates \cite{yan_pd_power}. Section~\ref{sec:limitations} shows how the stall depends on the idle level.

\subsection{Thermal Model Grounding and Hardware Cross-Check}
\label{sec:thermal_grounding}

We derive the boundary conditions and material parameters of the transient thermal model from Coenen et al.'s CFD- and FE-based characterization of 2.5D and 3D co-packaged optics \cite{imec_TCPMT}. The modeled stack follows the OIO3D configuration of Coenen et al., in which the XPU (the compute die, here a GPU) sits on top, followed by a hybrid-bonded EIC and an active photonic integrated circuit (aPIC) that hosts the modeled MRRs. The aPIC is bonded to a passive photonic interposer. Table~\ref{tab:coenen_params} lists every input of the model and its source. Coenen et al. report temperatures relative to the coolant inlet, and because the model is linear, the chosen coolant temperature shifts all temperatures equally and leaves the swing, and hence the stall, unchanged. The EIC and aPIC are constant background heat sources at their reported power densities, the XPU receives the phase-resolved power trace of Section~\ref{sec:experimental_setup}, and cooling is applied at the top surface of the XPU through a Si microchannel cold plate. The transient is solved with a fixed 1~ms time step over ten repetitions of the simulated iteration so that the trace reaches its periodic steady state, and the first 1.6~s, which contain the start-up transient, are discarded. Automatic time stepping is not used, because it coarsened to 15 to 175~ms during long idle intervals and quantized every derived stall to that step.

\textbf{Stack built from published layer values.} Several geometric and material inputs in Table~\ref{tab:coenen_params} are design choices of this work, so we repeated the transient with a stack built from published values wherever they exist. In the OIO3D configuration the EIC and aPIC substrates are removed \cite{imec_Thermal}. The aPIC is therefore resolved into a 0.22~$\mu$m silicon device layer that carries the monitored rings on a 2~$\mu$m buried oxide, the layer pair of a 300~mm silicon photonics platform \cite{imec_isipp}, under a 3~$\mu$m back-end oxide, and the EIC becomes a 10~$\mu$m device and wiring layer, these two thicknesses being chosen in this work. The oxides are assigned 1.38~W m$^{-1}$ K$^{-1}$, close to the bulk conductivity of fused silica of about 1.4~W m$^{-1}$ K$^{-1}$ near room temperature, with which the measured conductivity of buried oxide layers beneath a 0.22~$\mu$m silicon overlayer agrees \cite{goodson_box}. The XPU keeps the 775~$\mu$m thickness of a 300~mm wafer \cite{vanharen_wafer}, the passive interposer is the 100~$\mu$m TSV interposer of recent photonic interposer platforms, built on 300~mm wafers at AIM Photonics \cite{aim_interposer} and on 200~mm wafers at CEA-Leti \cite{leti_interposer}, and silicon is assigned 130~W m$^{-1}$ K$^{-1}$ at 330~K, interpolated from the recommended values of 133~W m$^{-1}$ K$^{-1}$ at 323~K and 119~W m$^{-1}$ K$^{-1}$ at 350~K \cite{ho_powell_liley}. With the heat sources, boundary conditions, and power schedules unchanged, the swing at the rings changes from 11.4 to 11.6~K for Mixtral and from 10.2 to 10.7~K for LLaMA-MoE, the mean temperature rises by about 1~K, and the delay per round changes from 27.7 to 27.5~ms and from 38.9 to 39.4~ms. Six of the eight injected per-type delays move by less than 2\%, and the largest change, in the LLaMA-MoE backward dispatch, is 18\%. The steepest cooling slope rises to 0.76 and 0.44~K/ms, because the thin device layer now carries little local heat capacity, while the slew rate that removes the stall remains 0.5~K/ms for both models. The stall is therefore governed by the GPU power transient and the cold plate rather than by the details of the photonic layers.

\begin{table}[t]
\centering
\small
\caption{Inputs of the transient thermal model. Values taken from Coenen et al.~\cite{imec_TCPMT} give the section, table, or figure of that paper; derived values give the relation used; values marked as chosen in this work are design assumptions of the modeled stack.}
\label{tab:coenen_params}
\renewcommand{\arraystretch}{1.12}
\begin{tabular}{|>{\raggedright\arraybackslash}p{5.9cm}|>{\raggedright\arraybackslash}p{3.6cm}|>{\raggedright\arraybackslash}p{4.1cm}|}
\hline
\textbf{Parameter} & \textbf{Value} & \textbf{Source} \\
\hline
\multicolumn{3}{|l|}{\textit{Geometry}} \\
\hline
Tile footprint (XPU, EIC, aPIC, interposer) & $25\times25$~mm$^2$ & \cite{imec_TCPMT}, Sec.~II \\
\hline
Hybrid and D2W bond thickness & 2.4~$\mu$m & \cite{imec_TCPMT}, Table I \\
\hline
XPU die thickness & 700~$\mu$m & chosen in this work \\
\hline
EIC thickness & 100~$\mu$m & chosen in this work \\
\hline
aPIC thickness & 50~$\mu$m & chosen in this work \\
\hline
Passive interposer thickness & 300~$\mu$m & chosen in this work \\
\hline
\multicolumn{3}{|l|}{\textit{Materials}} \\
\hline
Silicon (XPU, EIC, interposer) thermal conductivity & 150~W m$^{-1}$ K$^{-1}$ & near \cite{glassbrenner_si}, Table I (156 at 300~K) \\
\hline
Silicon density and specific heat & 2330~kg m$^{-3}$, 700~J kg$^{-1}$ K$^{-1}$ & \cite{crc_handbook} \\
\hline
Hybrid bond with vias (XPU to EIC, EIC to aPIC), effective conductivity & 5.0~W m$^{-1}$ K$^{-1}$ & derived, $t/R$ with $R=0.48$~mm$^2$ K W$^{-1}$ (zone with vias) \cite{imec_TCPMT}, Table I, applied over the whole interface \\
\hline
Bond without vias (aPIC to interposer), effective conductivity & 2.55~W m$^{-1}$ K$^{-1}$ & derived, $t/R$ with $R=0.94$~mm$^2$ K W$^{-1}$ (zone without vias) \cite{imec_TCPMT}, Table I, applied by us \\
\hline
Bond density and specific heat & 2500~kg m$^{-3}$, 800~J kg$^{-1}$ K$^{-1}$ & chosen in this work \\
\hline
aPIC effective conductivity (silicon devices in oxide) & 80~W m$^{-1}$ K$^{-1}$ & chosen in this work \\
\hline
aPIC density and specific heat & 2300~kg m$^{-3}$, 700~J kg$^{-1}$ K$^{-1}$ & chosen in this work \\
\hline
Copper (cold-plate mass sensitivity only) & 400~W m$^{-1}$ K$^{-1}$, 8960~kg m$^{-3}$, 385~J kg$^{-1}$ K$^{-1}$ & \cite{crc_handbook} \\
\hline
\multicolumn{3}{|l|}{\textit{Boundary conditions}} \\
\hline
Top-side heat-transfer coefficient (Si microchannel cold plate on the XPU) & $7\times10^4$~W m$^{-2}$ K$^{-1}$ & \cite{imec_TCPMT}, Sec.~III-B and Fig.~11 \\
\hline
Interposer-to-substrate resistance & 0.1~K W$^{-1}$ & \cite{imec_TCPMT}, Sec.~II, applied per tile by us \\
\hline
Equivalent bottom-side heat-transfer coefficient & $1.6\times10^4$~W m$^{-2}$ K$^{-1}$ & derived, $1/(R A)$ over the tile \\
\hline
Coolant temperature & 45~$^\circ$C & chosen in this work \\
\hline
\multicolumn{3}{|l|}{\textit{Heat sources}} \\
\hline
XPU power, GEMM phases (time-varying schedule) & 700~W & \cite{imec_TCPMT}, Secs.~II-B and IV-A \\
\hline
EIC power density, constant & 0.426~W mm$^{-2}$ (266~W) & \cite{imec_TCPMT}, Sec.~II-B \\
\hline
aPIC power density, constant & 0.113~W mm$^{-2}$ (71~W) & \cite{imec_TCPMT}, Sec.~II-B \\
\hline
\multicolumn{3}{|l|}{\textit{Numerics}} \\
\hline
Elements & 8-node thermal solids, $20\times20$ laterally (1.25~mm) & chosen in this work \\
\hline
Time step, length of the transient & 1~ms fixed, ten iterations, first 1.6~s discarded & chosen in this work \\
\hline
Monitored point & aPIC layer, tile center & chosen in this work \\
\hline
\end{tabular}
\end{table}

\textbf{Which device to model.} Devices in one expert-parallel group do not share a timeline, because the order and timing of their compute and communication differ. We build the power schedule for the device with the median compute share and for the device with the smallest one, and inject the larger of the two stall estimates for every round type, because an all-to-all round completes only when its slowest participant can transmit. With per-reticle ports the per-device load is nearly uniform, and the duty cycles of the two devices agree within 0.1\%, yet their stalls differ; for LLaMA-MoE the delay per round is 20~ms for the median device and 39~ms for the least-loaded one, because what matters is when each round falls relative to the cooling transients.

\begin{figure}[t]
\centering
\includegraphics[width=\linewidth]{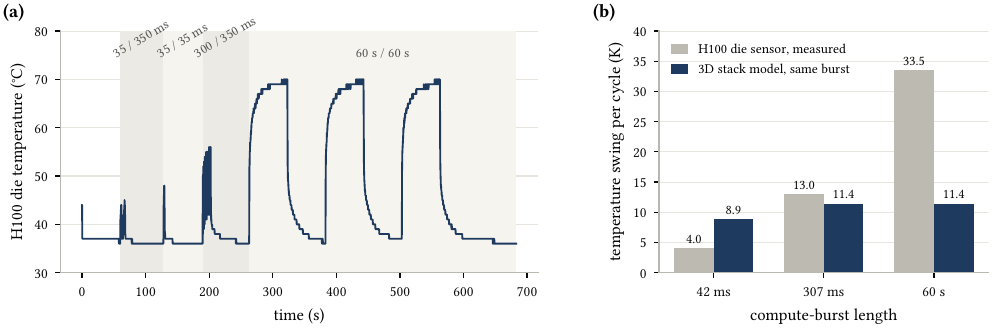}
\caption{Hardware cross-check on an NVIDIA H100 SXM GPU, read from the die-temperature sensor of the NVIDIA management library (NVML) every 20~ms. (a) Die temperature under four synthetic patterns, labeled by burst and idle length. (b) Per-cycle swing measured by the die sensor and computed for the 3D stack model under the same bursts between 131 and 692~W.}
\label{fig:calibration}
\end{figure}

\begin{figure}[t]
\centering
\includegraphics[width=\linewidth]{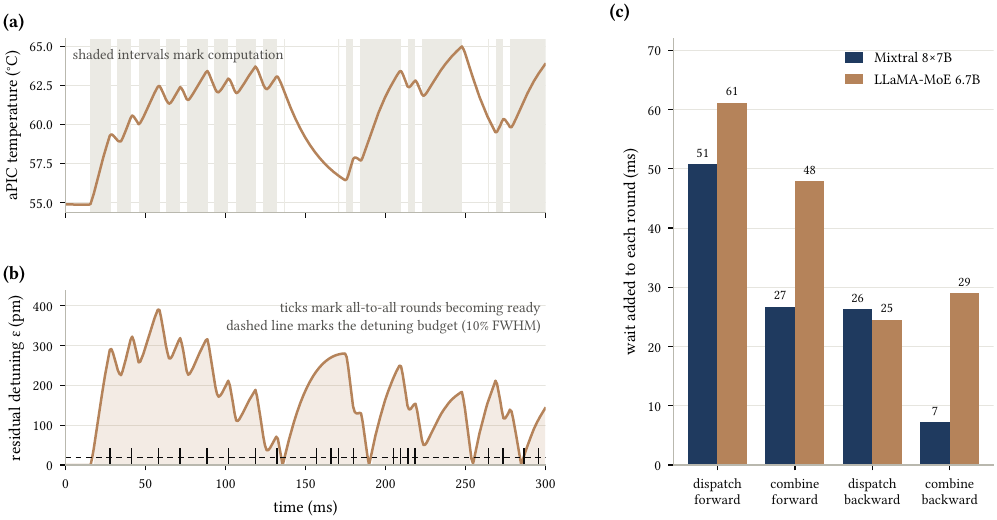}
\caption{From temperature to communication stall at the design point. (a) aPIC temperature over 300~ms of a LLaMA-MoE 6.7B iteration for the least-loaded device, with computation shaded. (b) Residual detuning $\varepsilon(t)$ after the 5~nm/s tracking loop, with the 19.4~pm budget (dashed) and the all-to-all rounds becoming ready (ticks). (c) Wait added to each round type, which is the delay injected into the network simulation.}
\label{fig:pic_mrr_transient}
\end{figure}

\textbf{Hardware cross-check.} No workload-driven temperature measurement of an OIO3D stack exists, so we compare the model's dynamics against a conventionally packaged GPU without fitting any model parameter to the measurement. On one H100 SXM node we ran synthetic burst and idle GEMM loops and a real Mixtral-style MoE training loop of forward and backward passes (eight GPUs, expert parallelism over NVLink, bf16), and we sampled the die-temperature sensor every 20~ms (Fig.~\ref{fig:calibration}). The sensor reports in 1~K steps. Under sustained GEMM load the H100 draws 692~W, which confirms the 700~W level of the power model, and it idles at 131~W. Two timescales appear. The die responds to 307~ms bursts with a 13~K per-cycle swing and to 42~ms bursts with about 4~K, so the fast swing is limited by how far heat diffuses during the burst. Under 60~s bursts the die swings by 33.5~K with a time constant of 6 to 7~s on heating and 9~s on cooling. NVML reports power averaged over a window longer than the short bursts, so we take the burst power as the sustained 692~W. For a direct comparison, we drive the thermal model of the stack, through a reduced-order model of two first-order lags fitted to its Ansys response (Section~\ref{sec:limitations}), with the same burst patterns between 131 and 692~W (Fig.~\ref{fig:calibration}(b)). For 307~ms bursts the model swings by 11.4~K against the measured 13~K, an agreement within 15\%. For 42~ms bursts the model swings by 8.9~K, more than the die sensor reports, as expected for a stack without a lid that is read at the photonic layer rather than through a die sensor with 1~K resolution. For 60~s bursts the model reaches 11.4~K and the measurement 33.5~K, and the difference is the slow component of the lidded package, which the directly cooled stack does not have. In the simulated models the compute bursts arrive back to back within each busy period, so the stack accumulates swings of 11.4 and 10.2~K (Fig.~\ref{fig:pic_mrr_transient}). In the same stack under representative MoE schedules, adding 3 or 10~mm of copper as cold-plate mass damps the swing by only 17 to 45\%, because the millisecond-scale swing is set by diffusion inside the die, and it raises the operating floor by up to 7.7~K. The measured MoE loop also shows that dispatch and combine move equal byte counts, with all-to-all phases occupying 44\% of a 1.59~s iteration on average (24 to 57\% across iterations).

\subsection{From Temperature to Communication Stall}
\label{sec:stall_derivation}

A temperature trajectory alone does not establish a communication outage. A fast cooling rate causes a stall only if the resulting resonance shift outruns what the tracking loop and the optical margin of the channel can absorb, so we convert $T(t)$ into per-round delays through an explicit chain.

First, the thermal shift of the ring resonance is $\Delta\lambda_{\text{th}}(t) = (d\lambda/dT)\,[T(t)-T(0)]$ with the silicon thermo-optic coefficient of 80~pm/K. Second, the control loop tracks this shift subject to its maximum wavelength slew rate $R_{\text{ctrl}}$. We adopt the measured tracking speed of the integrated drop-port locking loop of Agarwal et al.~\cite{PID}, which reports a tracking speed of 5~nm/s, a tuning range of 5~nm, and an accuracy better than 10~pm. The 5~nm/s is the fastest drift the loop was demonstrated to follow, and its slew is set by its step size and sampling rate at a cost in accuracy \cite{PID}. At 80~pm/K the tracking speed is 0.0625~K/ms (5~pm/ms). We report every stall for this loop, and Section~\ref{sec:mitigation} generalizes the result by treating both the slew rate and the loop bandwidth as design variables. Third, the residual detuning $\varepsilon(t) = |\Delta\lambda_{\text{th}}(t) - \lambda_{\text{track}}(t)|$ is the wavelength error the link actually sees. Fourth, the link is stalled while $\varepsilon(t)$ exceeds an acceptable-detuning budget $\varepsilon_{\text{max}}$ of 10\% of the resonance full width at half maximum (FWHM). For a drop filter this budget is conservative, since a Lorentzian loses 0.17~dB at that detuning and 1~dB at 25\% of the FWHM, while for a ring modulator, whose operating point sits on the slope of the resonance, a tenth of a linewidth is the scale at which the modulation amplitude degrades. Section~\ref{sec:limitations} shows that the stall depends only weakly on this choice up to 30\% of the FWHM. We take a loaded quality factor $Q$ of 8000, whose photon-lifetime bandwidth of 24~GHz supports 32~Gb/s signaling, so that the FWHM at 1550~nm is 193.8~pm and the budget $\varepsilon_{\text{max}}$ is 19.4~pm.

Finally, the stall is mapped onto communication. An all-to-all round that becomes ready while the ring is out of budget waits for the remainder of that stall window, and any stall that begins inside the round adds to its duration. This per-round delay, not a mean stall-window length, is what the network simulation injects before each round (Section~\ref{sec:e2e}), and Section~\ref{sec:closed_loop} feeds the stalled timeline back into the thermal model.

Fig.~\ref{fig:pic_mrr_transient} shows the chain at the design point. The aPIC swings by 11.4~K for Mixtral and 10.2~K for LLaMA-MoE between bursts and idle intervals, i.e., by 0.91 and 0.82~nm of resonance shift or more than four linewidths. The fastest cooling slopes reach 0.40 and 0.33~K/ms (median 0.09 and 0.12~K/ms), five to six times $R_{\text{ctrl}}$, and smaller than the 1.78~K/ms of a 0 to 100\% XPU power step \cite{imec_Thermal}. The residual therefore exceeds the budget for most of each busy block and returns inside it only briefly between bursts. What determines the communication penalty is where the all-to-all rounds fall relative to these windows. Dispatch rounds are issued immediately after the gating GEMM, at the start of the cooling transient, and combine rounds follow the expert GEMMs in the same way. The resulting delay per round, for the device class that stalls most, averages 28~ms for Mixtral and 39~ms for LLaMA-MoE. The per-type values we inject (dispatch and combine, forward and backward), taken as the larger of the two device classes, are 51, 27, 26, and 7~ms for Mixtral and 61, 48, 25, and 29~ms for LLaMA-MoE.

Two features of this result differ from a naive reading of the temperature traces. First, the device that stalls most is the one whose rounds fall at the start of its cooling transients. With nearly equal duty cycles, the phase of the rounds relative to the bursts, not the expert load, determines the stall. Second, the delay a round suffers is set by the stall window it lands in rather than by its own length. At 1.5~TB/s per GPU the rounds of Mixtral and LLaMA-MoE last 0.4 and 4.5~ms, far shorter than the 50 to 60~ms stall windows, so a round that becomes ready inside a window waits for nearly all of it.

\subsection{Network Baselines and the Design Point}
\label{sec:baseline_topo}

\begin{table}[t]
\caption{Stall-free iteration time in milliseconds for the simulated configurations of Table~\ref{tab:moe_config}, with dispatch rounds sized like combine rounds. Every fabric gives each GPU 1.5~TB/s of injection bandwidth, and the wafer fabric with CPO ports additionally provides the port bandwidth for off-wafer traffic.}
\label{tab:baseline_topo}
\centering
\small
\renewcommand{\arraystretch}{1.05}
\begin{tabular}{|l|c|c|}
\hline
\textbf{Fabric} & \textbf{Mixtral 8$\times$7B} & \textbf{LLaMA-MoE 6.7B} \\
\hline
Non-blocking fat-tree (ideal) & 841.3 & 427.4 \\
\hline
Direct all-to-all (``flat'') & 1152.7 & 1560.1 \\
\hline
Wafer, one 200~GB/s gateway link per wafer pair & 1954.9 & 493.1 \\
\hline
Wafer, 800~GB/s CPO port per reticle (design point) & \textbf{865.3} & \textbf{494.4} \\
\hline
\end{tabular}
\end{table}

Before the stall is quantified, the stall-free fabric is compared with conventional alternatives at equal per-GPU injection bandwidth, since comparisons at equal per-link rate favor whichever topology has more links per node. Table~\ref{tab:baseline_topo} gives the result. The connection between wafers dominates the result. With a single 200~GB/s optical gateway link between each pair of wafers, which funnels all traffic between two wafers through one reticle pair after extra intra-wafer hops, the wafer fabric takes 2.3 times as long as the ideal fat-tree for Mixtral, whose expert-parallel groups span two wafers. Giving every reticle its own CPO port, as Section~\ref{sec:wsc_sin} proposes, removes the funnel, and at 800~GB/s per port Mixtral's iteration time falls from 1955 to 865~ms, within 2.9\% of the ideal fat-tree. The 16-GPU expert-parallel group of LLaMA-MoE stays within one wafer, and LLaMA-MoE is 16\% slower than the fat-tree because of the two-hop relay paths of the row and column mesh.

The ports add 0.8~TB/s of off-wafer capacity per GPU on top of the 1.5~TB/s on-wafer injection, but this does not bias the comparison, because a fat-tree given the same 2.3~TB/s per GPU runs Mixtral and LLaMA-MoE in 839.3 and 426.3~ms, within 0.3\% of its 1.5~TB/s times. The flat fabric takes 1.33 and 3.2 times as long as the wafer fabric, because splitting 1.5~TB/s over up to 255 links leaves each link too narrow for the bursty rounds. The design rule that follows is a CPO port on every reticle rather than a shared gateway. With such ports an expert-parallel group may span wafers, as it does for Mixtral, whereas with a shared gateway only groups placed within one wafer, as for LLaMA-MoE, avoid the funnel. The wafer fabric thus approaches an ideal non-blocking fat-tree while needing far fewer cables and transceivers, which the next paragraph quantifies.

\textbf{Why wafer-scale.} In per-GPU terms, each reticle injects 1.5~TB/s into the wafer through twelve waveguides and a further 0.8~TB/s off the wafer through its CPO port. The on-wafer injection alone is 3.3 times the per-direction NVLink~4 bandwidth of an H100 SXM and 1.7 times the per-direction NVLink~5 bandwidth of a Blackwell GPU, whose bidirectional NVLink bandwidths are 900~GB/s and 1.8~TB/s \cite{nvlink}. On-wafer traffic needs no switch; it crosses one waveguide, or two with an OEO relay at a row and column intersection, and reaches every GPU of the wafer. Realizing 1.5~TB/s per GPU with a non-blocking fat-tree would take fifteen 800~Gb/s ports per GPU and, for the 256 GPUs of the Mixtral configuration built from 64-port 800~Gb/s switches, three tiers with 45 cables and 90 transceiver ends per GPU, whereas the wafer needs fourteen fibers per reticle for its port (seven per direction at 1.024~Tb/s each) and none for on-wafer traffic. This density is carried by the resonators, whose thermal behavior is the subject of the rest of this section.

\subsection{Tuning Stall without Thermal Feedback}
\label{sec:e2e}

The per-round delays of Section~\ref{sec:stall_derivation} are injected into the network simulation as a one-time stall before each all-to-all round of the corresponding type. They are not modeled as link propagation latency, which would compound on every packet, because a link recovers once its ring returns within the budget and the delay therefore applies once per round. In this open-loop estimate the delays are derived from the stall-free timeline and are not fed back. With the 5~nm/s loop the stall increases the iteration time by a factor of 1.58 for Mixtral 8$\times$7B (865 to 1371~ms) and 2.82 for LLaMA-MoE 6.7B (494 to 1396~ms). The ordering follows neither the temperature swing nor the communication volume. It follows the injected delay relative to the stall-free time. The iteration serializes the rounds of its four micro-batches, so the added time scales with the summed per-type delays of 111 and 163~ms per micro-batch, at about 4.5 and 5.5 times these sums. LLaMA-MoE combines the larger delays with the shorter iteration and therefore slows down more.

\subsection{Tuning Stall with Thermal Feedback}
\label{sec:closed_loop}

\begin{table}[t]
\caption{Iteration time with a 5~nm/s tuning loop, without and with the feedback of the stall onto the thermal field, in the simulated configuration and at full model depth.}
\label{tab:closed_loop}
\centering
\small
\renewcommand{\arraystretch}{1.1}
\begin{tabular}{|>{\raggedright\arraybackslash}p{5.6cm}|c|c|}
\hline
\textbf{Case} & \textbf{Mixtral 8$\times$7B} & \textbf{LLaMA-MoE 6.7B} \\
\hline
No tuning stall & 865~ms (1.00$\times$) & 494~ms (1.00$\times$) \\
\hline
Without feedback (open loop) & 1371~ms (1.58$\times$) & 1396~ms (2.82$\times$) \\
\hline
With feedback (closed loop) & 1140~ms (1.32$\times$) & 768~ms (1.55$\times$) \\
\hline
With feedback, full model depth (Section~\ref{sec:limitations}) & \textbf{about 1.16}$\times$ & \textbf{about 1.25}$\times$ \\
\hline
\end{tabular}
\end{table}

\begin{figure}[t]
\centering
\includegraphics[width=\linewidth]{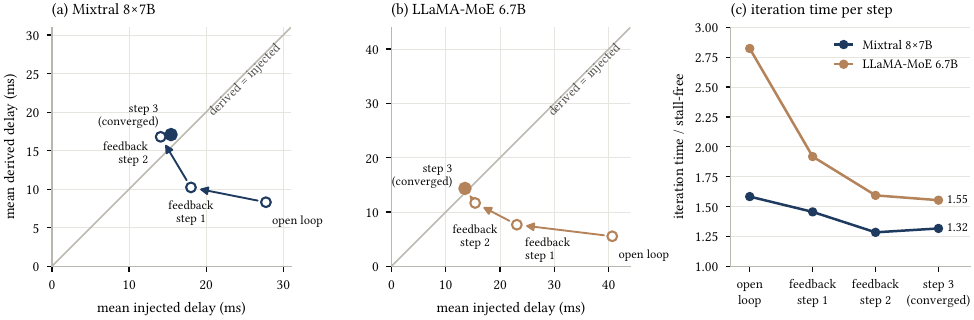}
\caption{Convergence of the coupled network and thermal simulation. Because the stall changes the GPU timeline and the timeline changes the temperature that sets the stall, the two simulations are run in turn. Each feedback step injects a delay map into the network simulation, recomputes the temperature from the resulting timeline, and derives a new map from it, so the steps belong to the calculation and not to the tuning of the ring. (a, b) Each point is one step, placed at the mean delay it injects and the mean delay it yields, both averaged over the four round types. The open loop starts from the stall-free timeline, each later step injects the average of the previously injected and derived maps, and step 3 (filled) lies on the line of equal delays, where the result is self-consistent. (c) Iteration time at each step relative to the stall-free iteration.}
\label{fig:closedloop}
\end{figure}

The open-loop delays come from a timeline that has no stall in it, whereas the stall changes the timeline. A stalled GPU waits longer, dissipates less heat, and swings less in temperature, which in turn shrinks the stall. In the hardware both effects settle together into one operating point, but the network simulation and the thermal simulation cannot be solved together, so we reach that point by running them in turn. In each feedback step a delay map is injected into the network simulation, the per-device compute schedules are rebuilt from the resulting timeline, the thermal transients are recomputed, and a new delay map is derived for the same devices. The operating point is reached when the derived map equals the injected one, and this self-consistent point is the physical result. The feedback steps are therefore steps of the calculation, not of the ring tuning. Table~\ref{tab:closed_loop} compares the self-consistent result with the open-loop estimate.

The feedback is strong (Fig.~\ref{fig:closedloop}). Injecting the open-loop stall stretches the iteration by factors of 1.58 and 2.82, so the duty cycle of the modeled devices falls from 51\% to 32\% (Mixtral) and from 41\% to 15\% (LLaMA-MoE), and the delays derived from this stretched timeline fall to 8.3 and 5.5~ms on average, 70 to 86\% below the open-loop values. Injected in turn, these shorten the iteration to 1006 and 605~ms, and the delays derived from that timeline swing back up to 21 and 26~ms, so feeding the derived delays back directly makes the result oscillate. We therefore inject at each feedback step the average of the previously injected and the newly derived delays. With this averaging the feedback converges toward the line on which the derived delay equals the injected one (Fig.~\ref{fig:closedloop}(a, b)), with the gap between the two falling from 7.8 to 2.7 to 1.6~ms for Mixtral and from 15.4 to 3.7 to 0.8~ms for LLaMA-MoE, and after three feedback steps they agree within 11\% and 6\%. At this self-consistent point the mean delay is 15 and 14~ms, the duty cycle of the modeled devices is 39\% and 27\%, and the tuning stall lengthens the iteration by factors of 1.32 and 1.55 (Fig.~\ref{fig:closedloop}(c)). At full model depth, the configuration of actual training, the corresponding factors are about 1.16 and 1.25 (Section~\ref{sec:limitations}). The closed-loop excess is 54\% and 30\% of the open-loop excess, which shows that the coupling between the network timeline and the thermal field is itself a first-order effect that the self-consistent analysis resolves. The stall cannot remove itself, because reducing it restores the duty cycle and with it the swing that causes it.

\subsection{Mitigation Requirements}
\label{sec:mitigation}

\begin{figure}[t]
\centering
\includegraphics[width=\linewidth]{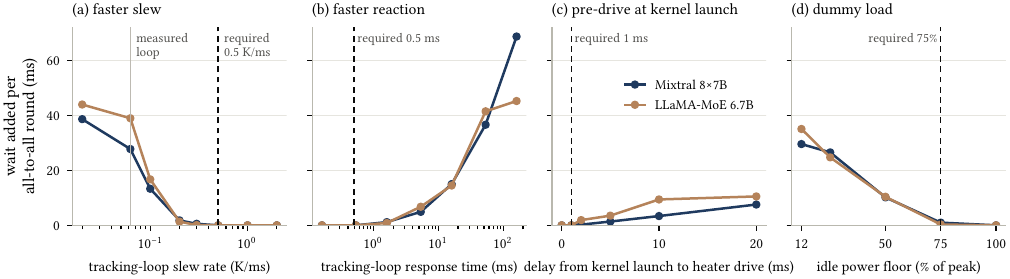}
\caption{Four ways to remove the stall and what each requires, on the open-loop traces. Each panel shows the mean wait added per all-to-all round as one design parameter varies, and the dashed line marks the value at which the wait falls to 1~ms or less. (a) Slew rate of the tracking loop; the solid grey line marks the measured 5~nm/s loop \cite{PID}. (b) Response time of a tracking loop without a slew limit; 0.5~ms corresponds to a loop bandwidth of 300~Hz. (c) Delay between a GPU kernel launch and the start of a predictive heater drive, at the measured slew rate. (d) Idle power floor of a dummy load held during waits, computed with the reduced-order thermal model.}
\label{fig:mitigation}
\end{figure}

Fig.~\ref{fig:mitigation} is a design chart with one panel per mitigation, from which a designer reads the requirement for a given implementation. All requirements in this subsection are evaluated on the open-loop traces, which have a higher duty cycle and swing than the closed loop of Section~\ref{sec:closed_loop} and therefore give upper bounds on the closed-loop requirements. The per-round delay falls to about 2~ms at 0.2~K/ms and to zero at 0.5~K/ms (40~nm/s), eight times the adopted value, for both models (Fig.~\ref{fig:mitigation}(a)). A slew limit is one of two ways a loop can fall behind. The other is a slow reaction, in which the loop can move the heater as fast as needed but notices the drift late (Fig.~\ref{fig:mitigation}(b)). We model this with a first-order tracking loop without a slew limit and characterize it by its response time $\tau$, the inverse of $2\pi$ times its bandwidth. During the steepest cooling the resonance moves by about 32~pm per millisecond, and such a loop trails it by roughly the drift accumulated over one response time, so the error stays inside the 19.4~pm budget only if $\tau$ is below about 0.6~ms. The simulation confirms this. The wait per round is about 15~ms for a 16~ms response time, 5 to 7~ms for 5~ms, about 1~ms for 1.6~ms, and zero for about 0.5~ms, which corresponds to a loop bandwidth of 300~Hz. Reported loops track at 5~nm/s with a 5~nm tuning range \cite{PID}, lock three rings over more than 40~$^\circ$C from a single monitor \cite{dong_array_lock}, and tune over more than 50~K at 3.8~$\mu$W/GHz with bit-statistics monitoring \cite{sun_bitstat}. A dithering loop has stabilized a ring against 10~Hz sinusoidal thermal fluctuations of magnitude 5~K \cite{padmaraju_dither}, a peak drift rate of 0.16 to 0.31~K/ms depending on whether the magnitude is read as peak to peak or as amplitude, which approaches the 0.5~K/ms requirement, and a thermal controller with a 500~Hz dominant pole has been reported \cite{mrr}. The requirement is therefore a specification for the EIC, and a modest one in loop bandwidth.

A second option is to pre-drive the heater at kernel launch. The GPU knows when each compute burst starts and ends, and the temperature follows from those events. If the GPU signals each kernel launch to the EIC, the controller can drive the heater along the temperature trajectory predicted by a thermal model and leave only the prediction error to the slow feedback loop. Fig.~\ref{fig:mitigation}(c) shows how the result depends on the delay between the kernel launch and the start of the heater drive. Within 1~ms the wait per round stays at zero, at 2~ms it is up to 2~ms, and at 10~ms it is 3 to 9~ms. Replaying the trajectory of the previous iteration instead cannot meet this, since all-to-all durations measured on the eight-GPU H100 system vary by tens of milliseconds between iterations.

Table~\ref{tab:mitigation} collects these options with two thermal ones and the drift-free ring of Section~\ref{sec:endurance}. Holding the GPU at a dummy load during its waits removes the swing at its source (Fig.~\ref{fig:mitigation}(d)). Raising the idle floor from 12\% to 50\% of peak power cuts the delay per round to about 10~ms, and 75\% brings it to 1~ms or less, at 130 to 160~W and 215 to 265~W of additional average power per GPU. Thermal mass damps the swing by at most 45\% and raises the operating point. The remaining option, a ring that does not drift, is examined in Section~\ref{sec:endurance}. The analysis thus gives each option a concrete specification.

\begin{table}[t]
\caption{Requirement and cost of each mitigation for a wait of 1~ms or less per all-to-all round for both models at the design point. Holding power is the continuous power the mitigation adds per GPU; the heater value is a lower bound at the most efficient heater reported for a closed-loop tuner (Section~\ref{sec:endurance}).}
\label{tab:mitigation}
\centering
\small
\renewcommand{\arraystretch}{1.1}
\begin{tabular}{|>{\raggedright\arraybackslash}p{3.3cm}|>{\raggedright\arraybackslash}p{3.0cm}|>{\raggedright\arraybackslash}p{2.3cm}|>{\raggedright\arraybackslash}p{2.6cm}|>{\raggedright\arraybackslash}p{2.4cm}|}
\hline
\textbf{Option} & \textbf{Requirement} & \textbf{Holding power per GPU} & \textbf{Added control} & \textbf{Tied to workload dynamics} \\
\hline
Faster slew & $\ge$0.5~K/ms (40~nm/s), eight times the demonstrated tracking speed \cite{PID} & $\ge$0.5~W heater bias & fast tracking loop & yes, steepest cooling slope \\
\hline
Faster reaction & response time about 0.5~ms (300~Hz) & $\ge$0.5~W heater bias & fast tracking loop & yes, steepest cooling slope \\
\hline
Heater pre-driven at GPU kernel launch & started within 1~ms of the compute burst & $\ge$0.5~W heater bias & kernel-launch signal to the EIC, thermal predictor & yes, burst timing \\
\hline
Dummy load during waits & idle floor $\ge$75\% of peak & 215 to 265~W & GPU power control & yes, idle intervals \\
\hline
Cold-plate thermal mass & not sufficient (swing reduced by 17 to 45\%) & none & none & not applicable \\
\hline
\textbf{Athermal LN ring, non-volatile setpoint} & athermal point within 3~K of the operating temperature & \textbf{none} (65~fJ per write) & \textbf{none} & \textbf{no} \\
\hline
\end{tabular}
\end{table}

\subsection{Removing the Drift at the Ring}
\label{sec:endurance}

Instead of tracking the drift faster, one can reduce it at the ring. Section~\ref{sec:background_ferro} introduced the two devices this requires, and here we test them against the thermal statistics of Section~\ref{sec:stall_derivation}.

\begin{figure}[t]
\centering
\includegraphics[width=\linewidth]{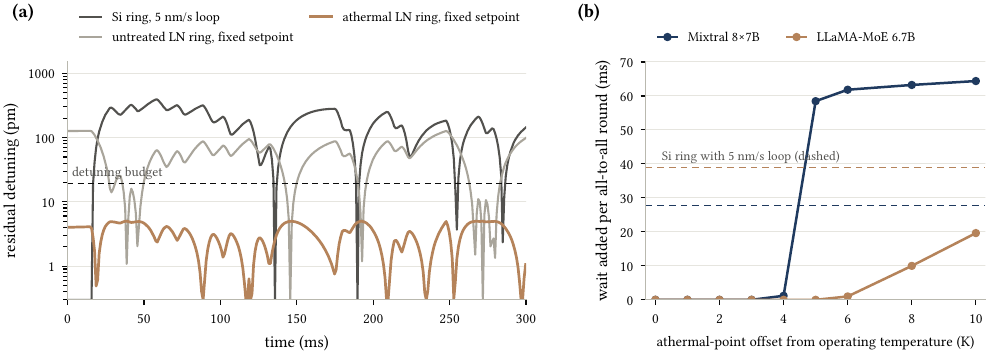}
\caption{Silicon ring versus lithium-niobate ring under the simulated workload. (a) Residual detuning during the LLaMA-MoE window of Fig.~\ref{fig:pic_mrr_transient}. The silicon ring drifts strongly and its loop follows too slowly, the untreated LN ring drifts less but is not followed at all, and the athermal LN ring barely drifts, so it needs no tracking loop. (b) Wait added per all-to-all round for an athermal LN ring with a fixed setpoint, versus the offset of its athermal point from the mean operating temperature.}
\label{fig:athermal}
\end{figure}

\textbf{Drift versus budget.} The steady-state aPIC swing is 11.4~K for Mixtral and 10.2~K for LLaMA-MoE. A silicon ring (80~pm/K) moves by 0.82 to 0.91~nm, four to five linewidths, and an untreated LN ring (18 to 30~pm/K, Section~\ref{sec:background_ferro}) by 184 to 342~pm, 10 to 18 times the 19.4~pm budget. For an athermalized LN ring the first-order term vanishes at the athermal point $T_a$ and the residual is $a\,(T-T_a)^2$ with $a=0.37$~pm/K$^2$ \cite{ln_athermal}. Fig.~\ref{fig:athermal}(a) applies the three cases to the same temperature trace. The silicon ring with the 5~nm/s loop exceeds the budget after every burst, the untreated LN ring with a fixed setpoint exceeds it for most of the window, and the athermal LN ring with a fixed setpoint, its athermal point at the mean operating temperature, stays within about 7~pm, inside the budget with no tracking loop at all. Fig.~\ref{fig:athermal}(b) turns this into the wait per all-to-all round. The athermal ring adds no wait while its athermal point lies within 3~K of the mean operating temperature for Mixtral (61~$^\circ$C) and within 5~K for LLaMA-MoE (60~$^\circ$C), compared with 28 and 39~ms per round for the silicon ring. Beyond that offset the fixed setpoint no longer covers the swing, and for Mixtral the wait exceeds that of the tracked silicon ring, so the offset tolerance is a design limit and the cladding has to be designed for the operating temperature of the stack. The mean operating temperature itself must therefore stay within about 3~K of the athermal point, which the coolant control of the stack has to hold; slow drifts of the mean within that window shift the residual by a few picometers and are absorbed by rewriting the non-volatile setpoint over seconds rather than milliseconds. The budget used here is that of a $Q=8000$ ring, and a ring with a narrower linewidth has a proportionally smaller budget and tolerance.

\textbf{Why a non-volatile element cannot track.} The ferroelectric setpoint cannot follow the drift of an untreated LN ring, because its range of 60 to 120~pm is smaller than the 184 to 342~pm of that drift. Even within its range, following the drift would take one write each time the resonance moved by one budget width, about $2\Delta\lambda/\varepsilon_{\text{max}}$ writes per compute burst over its heating and cooling legs, i.e., 19 to 35 writes per burst, or about 300 to 560 writes per iteration with 16 bursts per device per simulated iteration (four micro-batches, each with a forward and a backward pass of two compute blocks). The device has been cycled $10^7$ times \cite{ferroMRR}, as have HZO capacitors at 125~$^\circ$C \cite{hzo_endurance}, and $10^7$ writes at this rate last about $2\times10^4$ to $3\times10^4$ iterations. As a setpoint on an athermal ring, by contrast, the element is written at calibration and when the operating point changes, and even one write per 10~s over a $10^6$-iteration run at full depth is of order $10^6$ writes, within the demonstrated cycling.

\textbf{Power, range, and resolution.} A thermo-optic heater can only raise the ring temperature, so it must be biased above the coldest operating point by at least the full 0.9~nm swing, and every one of the 750 on-wafer endpoint rings of Section~\ref{sec:wsc_sin} and the 400 rings of the CPO port (200 wavelengths per direction) holds that bias continuously. At the 3.8~$\mu$W/GHz of the most efficient heater reported for an on-chip closed-loop tuner \cite{sun_bitstat}, covering the 0.91~nm (114~GHz) swing costs at least 0.43~mW per ring, or 0.5~W per GPU for its 1150 rings, and less efficient heaters cost proportionally more. A ring whose bias runs out must relock to the next free-spectral-range order, a mode hop that is itself a millisecond-scale stall. The non-volatile setpoint dissipates no holding power, and an update costs 65~fJ for a write and 358~fJ for a reset \cite{ferroMRR}. Its resolution fits the budget. The six states of one gate are 8 to 16~pm apart, about 12~pm on average, and with three gates the states after initialization are about 3~pm apart \cite{ferroMRR}, so a ring programmed to the nearest state sits within about 2 to 8~pm of its target, inside the 19.4~pm budget. A 15\% polarization loss corresponds to 9 to 18~pm, also inside the budget, and a rewrite restores it. Its range of 60 to 120~pm is the constraint. It covers the athermal residual of about 7~pm and a slow operating-point drift of tens of picometers, while an as-fabricated resonance offset larger than the range calls for a one-time trim outside it. Where a finer or faster correction is wanted, the Pockels effect of the same LN waveguide offers it at an efficiency of the order of the 7~pm/V reported for x-cut thin-film LN racetrack resonators \cite{ln_eo_ring}, so half a state corresponds to about 0.2 to 1.1~V across a capacitive load, within the supply range of an EIC.

\textbf{Comparison of the mitigations.} Table~\ref{tab:mitigation} places the athermal ring with a non-volatile setpoint beside the other options. Every option in the table except added thermal mass brings the wait per round to 1~ms or less, so the options differ in what they cost. The faster loop, the faster reaction, and the predictive heater drive all keep the silicon ring and its heater, and with it a holding power of at least 0.5~W per GPU, and they add either a tracking speed eight times the demonstrated one or a signal path from the GPU to the EIC with a thermal predictor. The dummy load needs no change to the ring but spends 215 to 265~W per GPU. Each of these four is also tied to the workload, since the loop must be designed for the steepest cooling slope, the prediction for the burst timing, and the dummy load for the idle intervals, so a different model or parallelization changes the requirement. The athermal ring with a ferroelectric setpoint is the only option that needs no holding power, no fast loop, and no signal from the GPU, and whose operation does not depend on the workload, because it removes the drift at its source rather than following it. Its requirement is on the device instead, namely an athermal point within about 3~K of the operating temperature and the integration of the TiO$_2$ cladding with the ferroelectric gate on one ring, which Fig.~\ref{fig:athermal} and the analysis above quantify.

\subsection{Sensitivity and Full Model Depth}
\label{sec:limitations}

\begin{figure}[t]
\centering
\includegraphics[width=\linewidth]{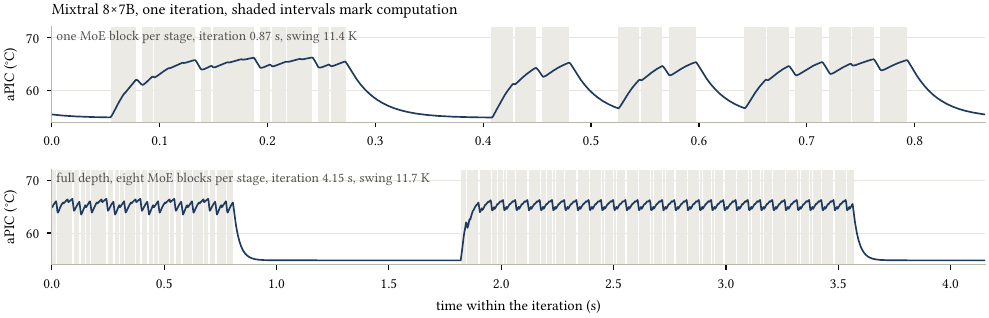}
\caption{aPIC temperature of Mixtral 8$\times$7B over one iteration with one MoE block per pipeline stage and with the eight blocks per stage of the full model, from Ansys, with computation shaded.}
\label{fig:fulldepth}
\end{figure}

\textbf{Full model depth.} In the full models each pipeline stage holds eight MoE blocks, so a device computes several blocks back to back before it waits on the other stages. Because the thermal model is linear, a reduced-order model of two first-order lags of 12 and 30~ms, with gains of 1.5 and 11.0~K at full power, fitted to one Ansys trace, reproduces the other traces within 0.3~K rms. Applied to full-depth schedules, it gives a wait of 15~ms per round for Mixtral and 11 to 16~ms for LLaMA-MoE, and a full Ansys transient of the Mixtral schedule over 41.5~s at 1~ms confirms it with a swing of 11.7~K and a wait of 14.7~ms per round (Fig.~\ref{fig:fulldepth}). The long stalls that follow a wait on the other stages then occur once per eight blocks instead of once per block. A round that becomes ready right after such a wait, while the stack heats from its cold state, waits 80 to 160~ms, and these rounds make up one forward round in four with one block per stage but only one in about thirty at full depth. The forward rounds therefore wait 5 to 9~ms instead of 27 to 51~ms, while the stalls set by the swing inside a busy period remain and the backward rounds wait 15 to 28~ms. With thermal feedback, the stall lengthens the training iteration at full model depth by a factor of about 1.16 for Mixtral and 1.25 for LLaMA-MoE. These factors scale the excess iteration time of the self-consistent point by the ratio of the full-depth to the single-block wait per round that the reduced-order model gives for the device class that stalls most, 15 to 30~ms for Mixtral and 16 to 35~ms for LLaMA-MoE.

\textbf{Sensitivity to the ring and power assumptions.} The stall depends on the quality factor $Q$ of the ring, which sets the width of its resonance, and on the detuning budget $\varepsilon_{\text{max}}$, the drift a link tolerates before it stalls. Re-evaluating the open-loop per-round delays over $Q\in[5000,15000]$ and $\varepsilon_{\text{max}}\in[5\%,20\%]$ of the FWHM moves the means from 28 to between 25 and 37~ms for Mixtral and from 39 to between 27 and 53~ms for LLaMA-MoE. Widening the budget alone to 25\% of the FWHM, the 1~dB point of a Lorentzian drop filter, gives 25 and 32~ms, and 30\% gives 25 and 27~ms. The stall depends only weakly on the budget because the detuning overshoots it several-fold. It is sensitive to the swing. Halving the temperature excursion, which matches the ratio between the measured and modeled swings for the 42~ms bursts of Section~\ref{sec:thermal_grounding}, gives 10 and 8~ms per round instead of 28 and 39~ms. Even at this lower swing the wait per round remains several times the 0.4 and 4.5~ms duration of the rounds themselves, so the stall stays a first-order term. The idle-power fraction matters less, since changing it from 6\% to 25\% of peak, which includes the measured 19\%, changes the delay by under 15\% for the median devices and by about a third for the least-loaded one. The loop is the other parameter the result hinges on, and Section~\ref{sec:mitigation} gives its full dependence.

\section{Related Work}
\label{sec:related_work}

\textbf{Wafer- and panel-scale optical interconnects.} Several efforts co-package photonics with compute and memory dies at wafer or panel scale to overcome the pin-density and reach limits of electrical I/O. TSMC's COUPE integrates photonic engines at the die level \cite{tsmc_coupe} and its system-on-wafer platform integrates full-reticle compute dies and memory on one wafer \cite{tsmc_sow}, Bao et al. analyze the cost and benefit of wafer-scale silicon photonics \cite{lightmatter}, Celestial AI reports an interposer-assembly process with optical bridges \cite{celestialai}, panel-scale reconfigurable photonic fabrics have been proposed \cite{panelscale}, 3D-integrated chip-to-chip photonic links have been demonstrated \cite{pic_3d}, and photonic device platforms for AI accelerators have been reviewed \cite{dwdm}. Bernadskiy et al. model MoE training on a 3D-integrated optical platform \cite{lightmatter_moe}. These works establish the architectural and bandwidth case for wafer- and panel-scale optics, largely under steady-state or synthetic traffic assumptions. This paper instead evaluates such a fabric under the workload-coupled thermal transients of real MoE training and shows that they materially affect the bandwidth these architectures deliver.

\textbf{Photonic tuning and thermal management.} The vulnerability of microring links to thermal drift is well documented at the device level, through wavelength-locking loop design \cite{PID, padmaraju_dither, sun_bitstat, dong_array_lock}, thermal characterization of co-packaged optics \cite{imec_Thermal, imec_TCPMT}, and thermally aware transmitter design \cite{mrr}, which quantify loop bandwidth and cooling behavior for individual rings or small arrays under controlled or steady-state loads. Non-volatile ferroelectric tuning of LN rings \cite{ferroMRR}, the switching speed \cite{hzo_switching_speed} and field-cycling behavior \cite{hfo2} of HfO$_2$-based ferroelectrics, and athermal lithium-niobate resonators \cite{ln_athermal, ln_athermal_broadband} have been characterized separately. None of these studies connects device-level thermal and tuning behavior to the communication timing of large-scale distributed training, which Sections~\ref{sec:thermal_grounding} to \ref{sec:closed_loop} do by deriving per-round stalls from a workload-driven thermal transient and feeding them back into the network timeline.

\textbf{Power-aware AI datacenters.} A growing line of work models and controls the power of AI fleets. Splitwise sizes prefill and decode pools under a power budget \cite{splitwise}, kernel-level estimators predict average GPU power from the executed kernels \cite{energaizer}, and analytical provisioning models relate per-GPU power to serving load \cite{yan_pd_power}. These models address the power of a deployment or of a whole workload, and Yan et al. model average rather than instantaneous power because short-term fluctuations of individual instances tend to average out across a large deployment \cite{yan_pd_power}. Co-packaged optics operate in the opposite regime. The rings sit within about a hundred micrometers of the compute die, so they follow the millisecond power swing of that one device, and averaging across devices does not remove it. This paper characterizes that per-device transient and its effect on communication, which complements fleet-level power models.

\textbf{Communication scheduling and reconfigurable fabrics for MoE.} At the network layer, MixNet reconfigures an optical-electrical fabric at runtime in response to MoE's skewed all-to-all traffic \cite{mixnet}, Chronos preschedules circuit-switched paths for the predictable traffic of LLM training and schedules MoE traffic on demand \cite{chronos}, Morphlux reconfigures a torus fabric for multi-tenant ML jobs \cite{morphlux}, InfiniteHBD builds a datacenter-scale high-bandwidth domain with optical circuit switching \cite{infinitehbd}, and Janus replaces the token all-to-all with scheduled expert fetching in MoE blocks where moving experts is cheaper than moving tokens \cite{ep}. The optical systems among these treat the reconfiguration or switching time of the optical layer as a fixed or assumed parameter rather than deriving it from the thermal behavior of the photonic devices, which is the gap our workload-to-thermal-to-optical-to-network methodology addresses.

This work is complementary to these lines and contributes the quantitative link between them. It quantifies, through a physically grounded and self-consistent chain from GPU power to ring detuning to communication stall, how much of a wafer-scale optical fabric's bandwidth a conventional thermo-optic tuning loop delivers under MoE training, and what a faster loop, a predictive heater drive, a dummy load, or a drift-free ring would each have to deliver to recover the rest.

\section{Conclusion}

This work analyzes wafer-scale photonic interconnects for MoE training with a cross-layer methodology. The methodology chains packet-level network simulation, a transient thermal model grounded in published stack data and cross-checked against H100 die-temperature telemetry, and an optical-domain stall criterion, and it feeds the resulting per-round stalls back into the network simulation. MoE execution produces 10 to 11~K temperature swings at the photonic layer between compute bursts and idle intervals. With a thermo-optic loop limited to the measured 5~nm/s, the stalls lengthen the training iteration of Mixtral 8$\times$7B and LLaMA-MoE 6.7B at full model depth by factors of about 1.16 and 1.25 once the stall is fed back, on a fabric whose stall-free iteration time is within 3\% and 16\% of that of an ideal non-blocking fat-tree. In the simulated configuration the factors are 1.32 and 1.55 with the feedback and 1.58 and 2.82 in an open-loop estimate. The thermal feedback reveals that the stall reshapes the duty cycle of each GPU and thereby its own thermal cause. The analysis further turns the penalty into specifications. A loop that slews at 40~nm/s or reacts within 0.5~ms, a heater pre-driven within about 1~ms of each GPU kernel launch, or a dummy load holding the GPUs above 75\% of peak power each removes it, and added thermal mass does not. Removing the drift at the ring is the least costly of these options. An athermalized lithium-niobate ring with a non-volatile ferroelectric setpoint keeps its residual inside the detuning budget if its athermal point lies within about 3~K of the operating temperature, and unlike the other options it needs no holding power, no fast loop, no signal from the GPU, and no tuning to the workload, whereas a ferroelectric actuator used to track the drift falls short on range and write endurance. For a wafer fabric provisioned well enough to approach a non-blocking network, the tracking capability of the tuning loop is a first-order requirement, together with a CPO port on every reticle.

\bibliographystyle{ACM-Reference-Format}
\bibliography{refs}

\end{document}